\documentclass[twocolumn,amsmath,amssymb,prb,floatfix]{revtex4-2}
\usepackage[usenames,dvipsnames,svgnames,table]{xcolor}
\usepackage{amsmath}
\usepackage{amsfonts}
\usepackage{bm}
\usepackage{dsfont}
\usepackage{graphicx}
\usepackage{subcaption}
\usepackage{braket}
\usepackage{booktabs}
\usepackage{multirow}

\newcommand{\nn}{\operatorname{NN}}

\DeclareMathOperator{\mlp}{mlp}

\usepackage{titlesec}

\usepackage[draft]{todonotes}
\usepackage[normalem]{ulem}

\begin{document}

\title{Machine learning of solvent effects on molecular spectra and reactions}

\author{Michael Gastegger}
\email{michael.gastegger@tu-berlin.de}
\affiliation{Machine Learning Group, Technische Universit\"at Berlin, 10587 Berlin, Germany}
\affiliation{ BASLEARN, BASF-TU joint Lab, Technische Universit\"at Berlin, 10587 Berlin, Germany}
\affiliation{DFG Cluster of Excellence ``Unifying Systems in Catalysis'' (UniSysCat), Technische Universit\"at Berlin, 10623 Berlin, Germany}
\author{Kristof T. Sch\"utt}
\email{kristof.schuett@tu-berlin.de}
\affiliation{Machine Learning Group, Technische Universit\"at Berlin, 10587 Berlin, Germany}
\author{Klaus-Robert M\"uller}
\affiliation{Machine Learning Group, Technische Universit\"at Berlin, 10587 Berlin, Germany}
\affiliation{Department of Artificial Intelligence, Korea University, Anam-dong, Seongbuk-gu, Seoul 02841, Korea}
\affiliation{Max-Planck-Institut f\"ur Informatik, Saarbr\"ucken, Germany}

\date{\today}

\begin{abstract}
Fast and accurate simulation of complex chemical systems in environments such as solutions is a long standing challenge in theoretical chemistry.
In recent years, machine learning has extended the boundaries of quantum chemistry by providing highly accurate and efficient surrogate models of electronic structure theory, which previously have been out of reach for conventional approaches.
Those models have long been restricted to closed molecular systems without accounting for environmental influences, such as external electric and magnetic fields or solvent effects.
Here, we introduce the deep neural network FieldSchNet for modeling the interaction of molecules with arbitrary external fields.
FieldSchNet offers access to a wealth of molecular response properties, enabling it to simulate a wide range of molecular spectra, such as infrared, Raman and nuclear magnetic resonance.
Beyond that, it is able to describe implicit and explicit molecular environments, operating as a polarizable continuum model for solvation or in a quantum mechanics / molecular mechanics setup.
We employ FieldSchNet to study the influence of solvent effects on molecular spectra and a Claisen rearrangement reaction.
Based on these results, we use FieldSchNet to design an external environment capable of lowering the activation barrier of the rearrangement reaction significantly, demonstrating promising venues for inverse chemical design.
\end{abstract}


\maketitle

\section{Introduction}
The presence of a solvent can dramatically change molecular properties as well as the outcome of reactions~\cite{varghese2019origins, reichardt2011solvents}.
Hence, a profound understanding of the interactions between molecules and their environments is tantamount not only for rationalizing experimental results, but also guiding the way towards controlling, or even designing, reactions and properties~\cite{varghese2019origins, zunger2018inverse, sanchez360}.
While computational chemistry has described such phenomena with reasonable success, obtaining accurate predictions which can be related to experiment is a highly non-trivial task.
Due to the large number of species involved, treating a system and its environment entirely with electronic structure theory is prohibitively expensive, especially for accurate high-level methods.
Approximate schemes, on the other hand, are often unable to capture important physical aspects, such as structural features of the environment in the case of continuum models or chemical reactions in the case of classical force-fields. 
To overcome these issues, we propose a deep neural network potential that includes the influence of external fields, to capture the interactions of the chemical system with the environment.

Recently, machine learning (ML) methods have emerged as a powerful strategy to overcome this trade-off between accuracy and efficiency inherent to computational chemistry approaches.\cite{von2020exploring, von2020retrospective,tkatchenko2020machine}
Highly efficient ML models now provide access not only to interatomic potential energy surfaces~\cite{behler2007generalized,braams2009permutationally,bartok2010gaussian,bartok2017machine,schutt2017quantum,smith2017ani,chmiela2018towards,podryabinkin2019accelerating,unke2019physnet}, but also to a growing range of molecular properties~\cite{rupp2012fast, montavon2013machine,schutt2018schnet,gilmer2017neural,faber2017prediction,shapeev2016moment,huang2020quantum}.
Specialized ML architectures have been developed for the prediction of vectorial and tensorial quantities~\cite{Thomas2018, anderson2019cormorant,grisafi2018symmetry}, such as dipole moments~\cite{gastegger2017machine}, polarizabilities~\cite{Wilkins2019} and non-adiabatic coupling vectors~\cite{westermayr2020combining}.
Such advances pave the way for using ML models in practical applications, with the simulation of infrared~\cite{gastegger2017machine}, Raman~\cite{raimbault2019using, sommers2020raman},  ultraviolet~\cite{zhang2020towards} and nuclear magnetic resonance (NMR) spectra~\cite{paruzzo2018chemical} being only a few examples.
At the same time, there is an ongoing effort to incorporate more physical knowledge into ML algorithms, giving rise to semi-empirical ML schemes~\cite{li2018tight, hegde2017machine} and even models based on electron densities~\cite{Ryczko2018, brockherde2017bypassing,bogojeski2019density} and wavefunctions~\cite{schutt2019unifying,carleo2017solving,hermann2020deep}.

Most of these approaches operate on closed systems, where the molecule is not subjected to any external environment.
Christensen \textit{et al.}~\cite{christensen2019operators} proposed a general framework for modeling response properties with kernel methods.
In this context, and in order to model electric field dependent properties, the FCHL representation~\cite{faber2018alchemical} was extended by an electric field model based on rough estimates of atomic partial charges.
This makes it possible to predict various response properties beyond atomic forces, such as dipole moments across compositional space as well as static infrared spectra.
Note that this approach inherently relies on molecular representations which are able to capture the required perturbations of the energy.

In this work, we propose the \textit{FieldSchNet} deep learning framework, which models the interactions with arbitrary environments in the form of vector fields.
As a consequence, our model is able to describe various implicit and explicit interactions with the molecular environment using external fields as a physically motivated interface.
Moreover, the field-based formalism automatically grants access to first and higher order response functions of the potential energy, with polarizabilities and nuclear magnetic shielding tensors being only a few examples.
This enables the prediction of a wide range of molecular spectra (e.g. infrared, Raman and NMR) without the need for additional, specialized ML models.

FieldSchNet can be used as a polarizable continuum model for solvation~\cite{mennucci2012polarizable} or to interact with an electrostatic field generated by explicit external charges in an QM/MM-like setup~\cite{senn2009qm} -- an approach we refer to as ML/MM.
As the model offers speed-ups of up to four orders of magnitude compared to the original electronic structure reference, we can employ FieldSchNet to investigate the influence of solvent effects on the Claisen rearrangement reaction of allyl-p-tolyl ether -- a study out of reach for conventional high-level electronic structure or ML approaches.
While these simulations would take 18~years with the original electronic structure reference, they could be performed within 5~hours with FieldSchNet.

FieldSchNet goes well beyond the scope of previous ML approaches by providing an analytic description of a chemical system in its environment.
We exploit this feature by designing an external field in order to minimize the height of the reaction barrier in the above Claisen reaction.
Hence, FieldSchNet constitutes a unified framework that enables not only the prediction of spectroscopic properties of molecules in solution, but even the inverse chemical design of catalytic environments.

\section{Results}

\subsection{FieldSchNet architecture}

\begin{figure*}
	\includegraphics[width=1.8\columnwidth]{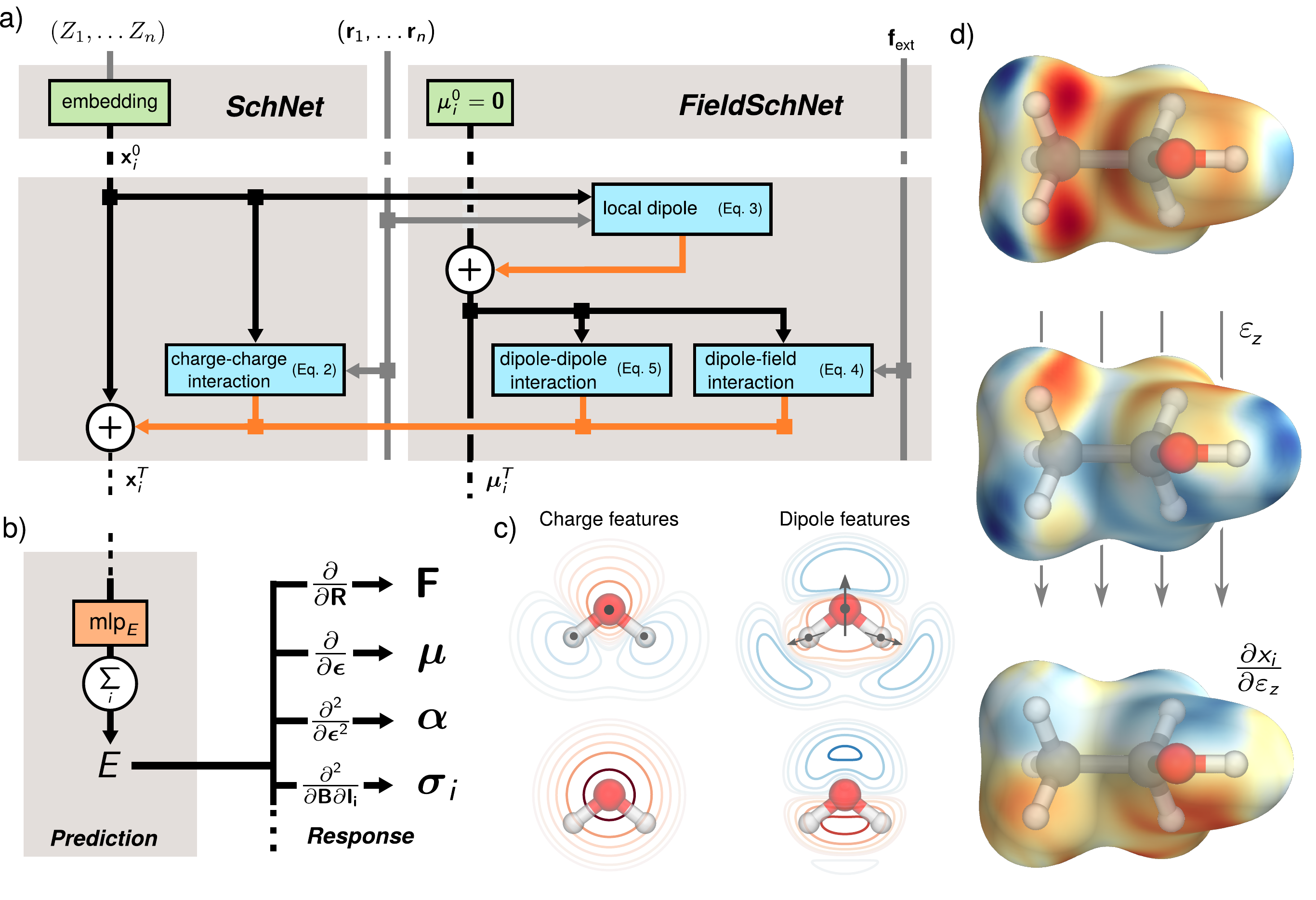}
	\caption{\textbf{Network architecture and learned representations.}
	(a) Scheme of the network blocks for generating the field-dependent representation. Starting from an initial representation of atom types and zero vectors for the dipole features, the conventional SchNet features $\mathbf{x}_i$ are augmented by a set of local dipole features $\boldsymbol{\mu}_i$. Interactions between the dipole features and the dipole features with the external field are used to update the representation $\mathbf{x}^T_i$ in each layer. 
	(b) The energies are predicted from the final set of features based on a sum of atomic contributions. By applying the appropriate derivative operations to the energy, different response properties can be computed, such as molecular forces $\mathbf{F}$, dipole moments $\boldsymbol{\mu}$, polarizabilities $\boldsymbol{\alpha}$ and shielding tensors $\boldsymbol{\sigma}_i$. 
	(c) Schematic comparison of charge and dipole based interactions in the water molecule shown for the whole molecule (top) and the oxygen atom (bottom). The introduction of dipoles allows for a more fine grained spatial resolution of the environment which is particularly clear to see in the case of the atomic oxygen contributions where charge like features only capture radial dependencies.
	(d) Representations for a hydrogen probe at position $\mathbf{R}$ projected onto a $\sum_i |\mathbf{R}_i - \mathbf{R}|^2$ isosurface, shown for the ethanol molecule. From top to bottom: FieldSchNet  descriptor without an external field, FieldSchNet descriptor in the presence of an external field applied to the z-plane of the molecule and response of the descriptor with respect to the z-component of an external field.
	}
	\label{fig:arch}
\end{figure*}

FieldSchNet allows to model interactions between atomistic systems and external fields $\boldsymbol{\epsilon}_\textrm{ext}(\mathbf{R})$ in a natural way.
It predicts molecular energies and properties based on local representations $\mathbf{x}_i \in \mathbb{R}^F$ of atomic environments \emph{embedded in external vector fields}, where $F$ is the number of features.
These representations are constructed iteratively, where the representation $\mathbf{x}_i^l$ of atom $i$ in each layer $l$ is refined via interactions with the neighboring atoms (see Fig.~\ref{fig:arch}a)
\begin{equation}
\mathbf{x}_i^{l+1} = \mathbf{x}_i^{l} + \mathbf{w}_i^{l} + \mathbf{u}_i^{l} + \mathbf{v}_i^{l}, \label{eq:schnet}
\end{equation}
starting from an initial embedding depending only on the respective atom type.
Here, $\mathbf{w}_i^{l}$ is the standard SchNet interaction (Fig.~\ref{fig:arch}a, left block), while the added terms $\mathbf{u}_i^{l}$ and $\mathbf{v}_i^{l}$ correspond to dipole-field and dipole-dipole interactions (Fig.~\ref{fig:arch}a, right block).
The SchNet interaction update then takes the form
\begin{equation}
\mathbf{w}_i^{l+1} = \nn\left[ \sum_{j}^{N} \mathbf{x}_j^{l} \mathbf{W}_q^{l}(r_{ij}) \right]. \label{eq:oschnet}
\end{equation}
The radial interaction functions $\mathbf{W}_q^{l}$ depend on the interatomic distance $r_{ij}$ and are learned from reference data.
A fully-connected neural network $\nn$ is applied afterwards performing a non-linear transformation.

In terms of rotational symmetry, the SchNet feature refinements can be interpreted as charge-charge interactions, as the features of $\mathbf{x}_i$ are scalars.
Hence, the internal structure of SchNet, as well as the generated representation, is invariant with respect to rotations and translations of the molecule.
This is an important requirement for machine learning potentials, as the energy of an atomic system exhibits the same symmetry~\cite{schutt2020learning}.
However, this invariance breaks down in the presence of external fields.
In this case, models also need to be constructed such that they are able to resolve rotations and translations relative to an external frame of reference.

FieldSchNet achieves this requirement by introducing an additional vector valued representation based on atomic dipole moments $\boldsymbol{\mu}_i \in \mathbb{R}^{F \times 3}$ (left side of Fig~\ref{fig:arch}a), which is refined  analogous to Eq.~\ref{eq:schnet}:
\begin{equation}
\boldsymbol{\mu}_i^{l+1} = \boldsymbol{\mu}_i^l + \sum_j \nn_q(\mathbf{x}_j^l) \mathbf{R}_{ij} \text{f}_\text{cutoff}(r_{ij}).\label{eq:dipfeat}
\end{equation}
Here, $\mathbf{R}_{ij}$ is the vector pointing from atom $j$ to $i$ and $f_\mathrm{cutoff}$ is a cutoff function constraining the influence of neighbors to the local environment.
The expression in Eq.~\ref{eq:dipfeat} generates a set of local vector-valued features on each atom, the orientation and magnitude of which is modulated by the surrounding atoms.

Based on these features, FieldSchNet models the interaction between molecule and external fields $\mathbf{u}_i^l$ with the term
\begin{equation}
\mathbf{u}_i^l = \nn_\epsilon\left[ (\boldsymbol{\mu}_i^l)^\top \boldsymbol{\epsilon}_\text{ext}(\mathbf{R}_i)\right],
\end{equation}
where $\boldsymbol{\epsilon}_\text{ext}(\mathbf{R}_i)$ is the field acting on atom $i$.
The dot product corresponds to the physical expression for first order approximation to the energy of a dipole in an external field.
This ensures that the orientation with respect to the external field is captured correctly and reduces to a constant in the absence of a field.

In addition to the dipole-field interaction, FieldSchNet introduces an update $\mathbf{v}_i^l$ based on the interaction between neighboring dipoles 
\begin{equation}
\mathbf{v}_i^l = \nn_\mu \left[\sum_j (\boldsymbol{\mu}_i^l)^\top \mathbf{T}^l(\mathbf{R}_{ij}) \boldsymbol{\mu}_j^l \right],
\end{equation}
where $\mathbf{T}(\mathbf{R}_{ij})$ is a 3-by-3 Cartesian interaction tensor inspired by the classical dipole-dipole interaction
\begin{equation}
\mathbf{T}(\mathbf{R}_{ij}) = \frac{3 r_{ij}^2 \mathbf{I} - \mathbf{R}_{ij}\mathbf{R}_{ij}^\top}{r_{ij}^5} \mathbf{W}_{\mu}^{l}(r_{ij}),
\end{equation}
which guarantees the right geometrical behavior. 
Similar to the SchNet interaction in Eq.~\ref{eq:oschnet}, $\mathbf{W}(r_{ij})_\mu^l$ is a learnable radial filter allowing the network to modulate the interaction strength.

For each external field, FieldSchNet adds a set of dipole features and expands the update in Eq.~\ref{eq:schnet} by the corresponding dipole-field and dipole-dipole terms.
As the scalar features $\mathbf{x}_i$ of the next layer depend on the dipole and field interactions of the current layer, those in turn are coupled with the preceding scalar features via the dipoles (Eq.~\ref{eq:dipfeat}).
This allows FieldSchNet to capture interactions with external fields far beyond the linear regime.

As visualized in Fig.~\ref{fig:arch}d at the example of an ethanol molecule, the FieldSchNet descriptor exhibits the same symmetry as the molecule (top).
Upon introducing an electric field in the z-axis of the molecule, the descriptor adapts to the changed environment and the original symmetry is broken (middle).
This effect can also be observed in the response of the descriptor with respect to the field (bottom).
At the same time, multiple successive SchNet and dipole-dipole updates enable the efficient construction of higher-order features of the molecular structure (see Fig.~\ref{fig:arch}c), going beyond the purely radial dependence of charge-like features.
Although inspired by electric dipoles $\boldsymbol{\mu}$, the dipole-like representations in FieldSchNet are auxiliary constructs and may also represent other quantities, e.g. the nuclear magnetic moments $\mathbf{I}_i$ when modeling magnetic fields.
An appropriate form of the dipoles corresponding to each field is learned purely data-driven.

Once the FieldSchNet representation $\mathbf{x}_i$ has been constructed, the potential energy is predicted via the atomic energy contributions typical for atomistic networks (Fig.~\ref{fig:arch}b)
\begin{equation}
 E\left(\mathbf{R}, \boldsymbol{Z}, \boldsymbol{\epsilon}_\text{ext}^{(\alpha)}, \ldots, \boldsymbol{\mu}_{0,i}^{(\alpha)}, \ldots \right) = \sum_i \nn_\text{E}(\mathbf{x}_i).
\end{equation}
Since the features $\mathbf{x}_i$ depend on the interactions between dipoles and fields of each previous layer and the dipoles are in turn constructed from the features, the potential energy is now a function of the atomic positions, nuclear charges and all external fields as well as initial atomic dipoles.
This makes it possible to obtain response properties from the energy model by taking the corresponding derivatives.

\subsection{Molecular spectroscopy}\label{sec:response}
FieldSchNet is particular promising for the prediction of molecular spectra, as a \emph{single} model provides access to a wide range of response properties. 
A variety of spectra can be simulated in this manner, ranging from vibrational spectra such as infrared and Raman to nuclear magnetic resonance spectra.
The availability of molecular forces makes it possible to go beyond static approximations and even derive vibrational spectra from molecular dynamics simulations~\cite{thomas2013computing}.
Moreover, the high computational efficiency of the machine learning model renders otherwise costly path integral molecular dynamics simulations feasible, which are able to account for nuclear quantum effects~\cite{sauceda2020dynamical} and yield predicted spectra close to experiment~\cite{habershon2013ring}.

We train a single FieldSchNet model on reference data generated with the PBE0 functional for an ethanol molecule in vacuum.
A combined loss function incorporates the energy ($E$), atomic forces ($\mathbf{F}$), dipole moment ($\boldsymbol{\mu}$), polarizability ($\boldsymbol{\alpha}$) and nuclear shielding tensors ($\boldsymbol{\sigma}$) as target properties (see Supplementary Text~1 for details).
Excellent fits were obtained for all quantities, as can be seen based on the test accuracy reported in Supplementary Tab.~3.
Energies and force predictions fall well within chemical accuracy, with mean absolute errors (MAEs) of 0.017~kcal/mol and 0.128~kcal/mol/\AA.
The response properties of the electric field exhibit MAEs as low as 0.04~D ($\boldsymbol{\mu}$) and 0.008~Bohr$^3$ ($\boldsymbol{\alpha}$) compared to value ranges of 4.56~D and 13.58~Bohr$^3$ present in the reference data.
In a similar manner, FieldSchNet yields low MAEs for the shielding tensor $\boldsymbol{\sigma}$, exhibiting an error of 0.123~ppm for hydrogen (range of 29.43~ppm) and 0.194~ppm for carbon atoms (range of 153.94~ppm).

\begin{figure*}
	\includegraphics[width=1.9\columnwidth]{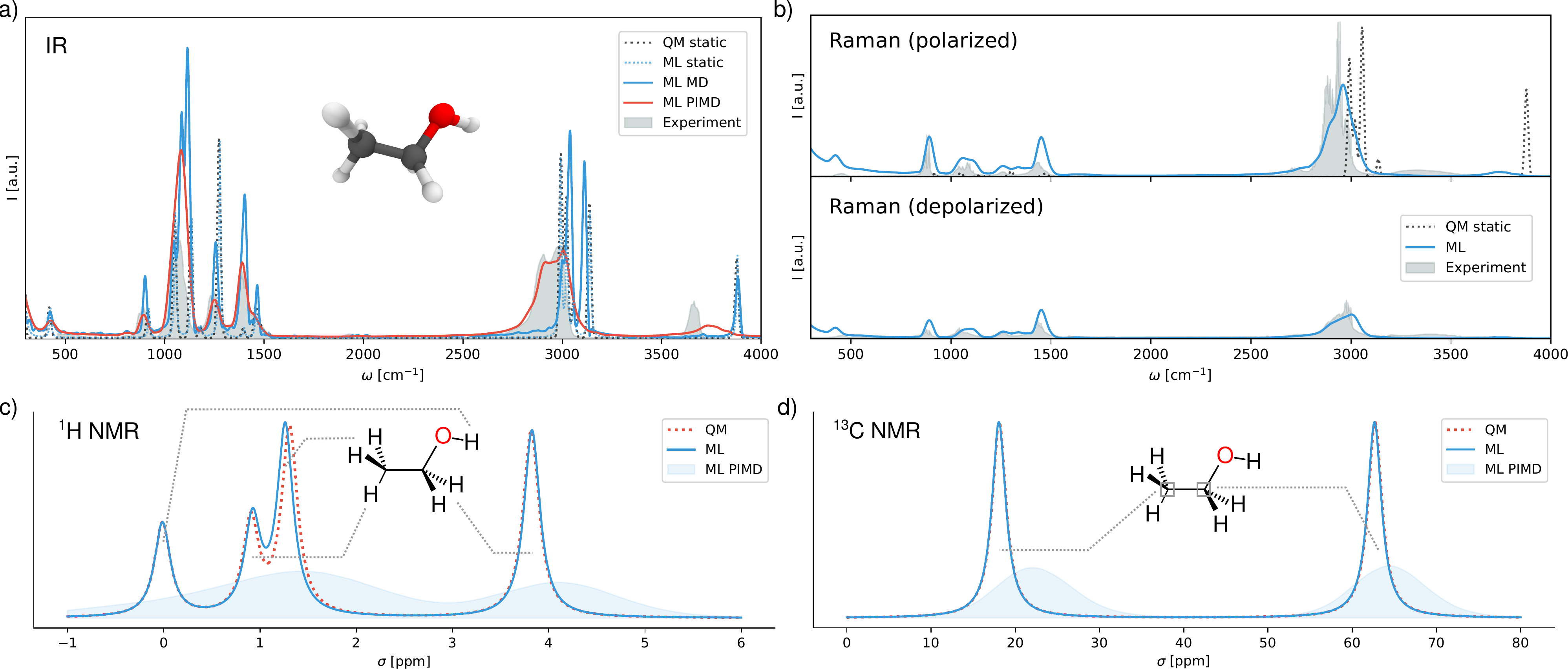}
	\caption{\textbf{Predicted spectra}(a) Infrared spectra for ethanol in vacuum predicted by the different methods compared to an experimental spectrum. (b) Polarized and depolarized Raman spectra simulated with FieldSchNet. The experimental spectra are shown in gray. (c) and (d) $^1$H and $^{13}$C chemical shifts predicted by the model and reference method, as well as the shift distributions sampled during path integral molecular dynamics.}
	\label{fig:spectra}
\end{figure*}

In the following, we simulate a range of spectra using the multi-property FieldSchNet model.
Fig.~\ref{fig:spectra}a shows infrared spectra obtained from static calculations and the dipole time-autocorrelation functions simulated by molecular dynamics.
In addition, a static spectrum obtained with the reference method, as well as the gas-phase experimental spectrum are provided\cite{NIST}.
While the FieldSchNet model is able to reproduce the static reference almost exactly, comparison of both spectra to experiment demonstrates the drawbacks of relying on purely static calculations for the prediction of vibrational spectra in general.

The potential of the machine learning model becomes apparent when going beyond the static picture.
In order to predict vibrational spectra from molecular dynamics simulations, a large number of successive computations are necessary, which quickly become prohibitive when relying on electronic structure methods.
However, due to the computational efficiency of FieldSchNet, these simulations can be carried out with little effort.
A single evaluation of all response properties takes 220~ms on a Nvidia Tesla P100 GPU compared to a computation time of 207~s with the original electronic structure method on a Intel Xeon E5-2690 CPU, a speedup by almost three orders of magnitude.
As a consequence, a simulation which would take 240~days with conventional approaches can now be performed in approximately 6~hours.
This huge step in efficiency grows even stronger for larger systems.

The benefits of performing molecular dynamics simulations can be observed in the spectrum recovered in this manner, as it exhibits a much better agreement with experiment in the low frequency regions.
Nevertheless, this approach still fails to reproduce positions and intensities of the bands associated with the stretching vibrations of the C-H bonds ($\sim$ 3000~cm$^{-1}$) and the O-H bond ($\sim$ 3600~cm$^{-1}$).
These differences are primarily due to the neglect of anharmonic and nuclear quantum effects.
One way to include these effects is via path integral molecular dynamics, where multiple coupled replicas of the molecule are simulated.
While this approach is computationally more demanding than classical molecular dynamics due to the additional replicas, the simulations can still be carried out efficiently with FieldSchNet.
As can be seen in Fig.~\ref{fig:spectra}a, accounting for anharmonic effects does indeed shift the C-H stretching vibrations to the experimental wavelengths and improves the position the O-H band, yielding a predicted spectrum close to experiment.
The remaining difference in the O-H band can most likely be attributed to a general limitation of the reference method.

In addition to infrared spectra, FieldSchNet enables the simulation of Raman spectra, which rely on molecular polarizabilities $\boldsymbol{\alpha}$.
Since the full polarizability tensor is predicted by the model, it is possible to compute polarized as well as depolarized Raman spectra.
Fig.~\ref{fig:spectra}b depicts both types of spectra as obtained with FieldSchNet via path integral molecular dynamics, as well as their experimental counterparts~\cite{kiefer2017simultaneous}.
In both cases, very good agreement with experiment is observed, with the most prominent difference being once again the vibrations in the O-H stretching regions.
The quality of the predicted spectra is particularly striking, when comparing to the statically computed polarized Raman spectrum, which fails to reproduce the shapes and magnitudes of several peaks.

Beyond vibrational spectra, FieldSchNet can be used to obtain NMR chemical shifts via the nuclear shielding tensors $\boldsymbol{\sigma}_i$.
In this manner, chemical shifts can be obtained for all NMR active isotopes, which in the case of ethanol are $^1$H, $^{13}$C and $^{17}$O.
The predicted and reference chemical shifts for the equilibrium configuration of ethanol are provided in Fig.~\ref{fig:spectra}c and d.
In addition, the distribution of shifts sampled during path integral molecular dynamics are shown.
$^{17}$O shifts are omitted from the analysis, as only one oxygen nucleus is present.
However, the shifts are still reproduced accurately and the associated error is given in Supplementary Tab.~3.
FieldSchNet predictions agree closely with the reference method for the $^1$H and $^{13}$C isotopes.
The $^1$H chemical shifts of the hydrogens in the CH$_2$ and CH$_3$ groups are close to their expected experimental values of 1.2 ppm and 3.8 ppm, respectively.
The peak shifted to 1~ppm is associated with the hydrogen atom in plane with the O-H bond. 
The resulting band structure vanishes in the molecular dynamics simulation due to rotations of the methyl group.
A major disagreement with experiment is the shift of the hydrogen in the O-H group which shows uncharacteristically low values of 1~ppm.
This can be attributed to a shortcoming of the reference method, which exhibits the same behavior.
The $^{13}$C shifts of the CH$_2$ and CH$_3$ carbon atoms agree almost perfectly with their experimental values of $\sim$60~ppm and $\sim$20~ppm.

\subsection{Implicit environments with polarizable continuum models}

Accounting for effects of the molecular environment and solvent effects in particular is crucial for a wide range of chemical applications.
These effects can critically influence the properties of compounds and the outcome of chemical reactions.
Due to the large number of species involved, treating a molecule and surrounding environment entirely with electronic structure methods is impractical.
In the case of solutions, approximating the solvent by a polarizable continuum model (PCM) with specific dielectric constant $\varepsilon$ has proven as a powerful tool~\cite{mennucci2012polarizable}.

By using an expression for the external field adapted from the reaction field approach of Onsager~\cite{onsager1936electric}, FieldSchNet can operate as a machine learning model for polarizable continuum solvents with an explicit dependency on $\varepsilon$ (see Sec~\ref{asec:pcm}).
To study this mode of operation, we train such a polarizable continuum FieldSchNet (pc-FieldSchNet) on a reference data set composed of computations for a ethanol molecule in the gas phase, as well as ethanol ($\varepsilon=24.3$) and water ($\varepsilon=80.4$) continuum solvents.
Again, the model is trained to predict the potential energy, the atomic forces and the response properties for the molecular spectra.
The polarizable continuum FieldSchNet (pc-FieldSchNet) reproduces all quantities with high accuracy, comparable with that of the model for response properties in vacuum (see Supplementary Tab.~3).
This demonstrates that pc-FieldSchNet is able to learn the correct dependence on the specific dielectric constant $\varepsilon$.

Supplementary Tab.~3 furthermore shows results for two benchmark datasets of methanol ($\varepsilon=32.63$) and toluene ($\varepsilon=10.3$), solvents that have not been included in training.
We find that pc-FieldSchNet generalizes well to these solvents with errors comparable to the prediction of solutions used for training.
The accuracy for toluene is slightly lower with the polarizability showing particularly high mean absolute errors of 0.243~Bohr$^3$.
However, this is can be attributed to an insufficient sampling of the regions of low polarity, as the most similar solvent included in training is vacuum.
The methanol dataset is reproduced with high accuracy, demonstrating that the model is able to generalize across unseen continuum solvents.

\subsection{Explicit environments with ML/MM}

Although continuum models are powerful tools for the description of solutions, they break down in situations where direct interactions between molecule and environment need to be considered, e.g. solute-solvent hydrogen bonds.
In these cases, the solvent has to be treated explicitly, e.g. using QM/MM schemes~\cite{cramer2013essentials, senn2009qm}.
These retain the full atomistic structure of the environment but instead treat it with more affordable classical force fields, while the molecule itself is described with electronic structure methods.
By expressing the external electric field as the field generated by the point charges of the MM region, FieldSchNet can operate in a similar manner and replace the quantum region in such a simulation, essentially yielding an ML/MM approach.
We use generalized atomic polar tensor charges~\cite{cioslowski1989general} computed with FieldSchNet to model the electrostatic interactions between both regions (see Sec~\ref{asec:mlmm} for details).
In the following, we study the impact of using an ML/MM solvent model on the simulated infrared spectrum of liquid ethanol.

We train a FieldSchNet model on a set of PBE0 reference data of ethanol configurations polarized by external charge distributions that have been sampled from a classical force field.
The test set performance of the resulting model is provided in Supplementary Tab.~3.
We observe slightly increased errors for the ML/MM model in energy and forces compared to the vacuum and continuum models due to the more complex environment.
Still, the model is able to reach high accuracy.

\begin{figure}
	\includegraphics[width=\columnwidth]{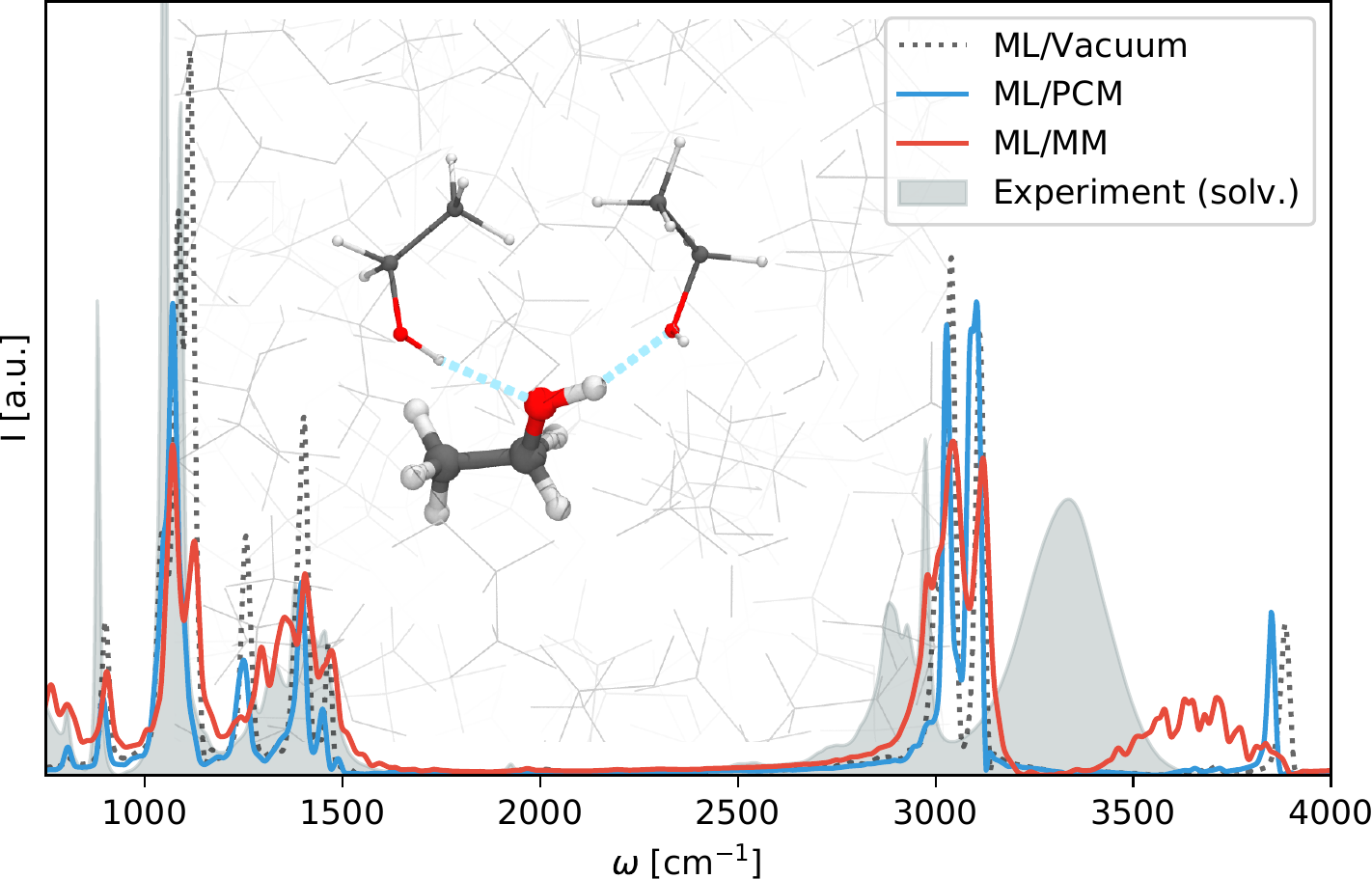}
	\caption{\textbf{Solvent effects on the infrared spectrum of ethanol.} Infrared spectra of liquid ethanol predicted without solvent (ML/Vacuum), with continuum solvent (ML/PCM) and explicit solvent molecules (ML/MM). The experimental spectrum is shown in gray. The inset depict hydrogen bonds between the hydroxyl group and the environment, responsible for the broadening and red-shift of the O-H stretching vibration.}
	\label{fig:ethqmmm}
\end{figure}

ML/MM molecular dynamics simulations are carried out for a single machine learning ethanol surrounded by a solvent box of 1250 ethanol molecules treated at force field level.
The resulting infrared spectrum is depicted in Fig.~\ref{fig:ethqmmm}, alongside spectra computed in the same manner using the vacuum model (Sec.~\ref{sec:response}) and continuum model ($\varepsilon=24.3$), as well as an experimental spectrum of liquid ethanol\cite{doroshenko2012infrared}.
Comparing the gas phase and continuum models, we find that in this case implicit solvent effects lead to no major improvements with respect to experiment.
The spectrum simulated via the ML/MM approach on the other hand, yields significantly better predictions.
The low wavelength regions in particular show excellent agreement with experiment.
The high wavelength regions are shifted to higher wavelengths since anharmonic effects are neglected in the classical ML/MM simulation.
However, we still observe the broadening and red-shift of the O-H stretching vibration present in the experimental spectrum.
This effect is caused by hydrogen bonding between the O-H groups in the machine learning region and the surrounding ethanols (see inset Fig.~\ref{fig:ethqmmm}).
Continuum models fail to account for these kind of interactions, as they neglect the structure of the solvent.

\subsection{Modeling organic reactions}
Effects of the environment and solvents play a central role in many molecular reactions.
Certain arrangements and combinations of molecules in the environment can promote or inhibit reactions in a dramatic fashion.
As such, a good understanding of these effects is crucial for the development of new catalysts and drugs.
However, accurate computational simulations of such systems are hard to obtain since intense sampling procedures are required in order to obtain reliable free energy profiles of the studied reaction.
This is further complicated by the need to account for the large number of molecules in the environment, which in many cases cannot be described by more affordable continuum models of solvation due to explicit interactions between solute and environment.

\begin{figure*}
	\includegraphics[width=1.8\columnwidth]{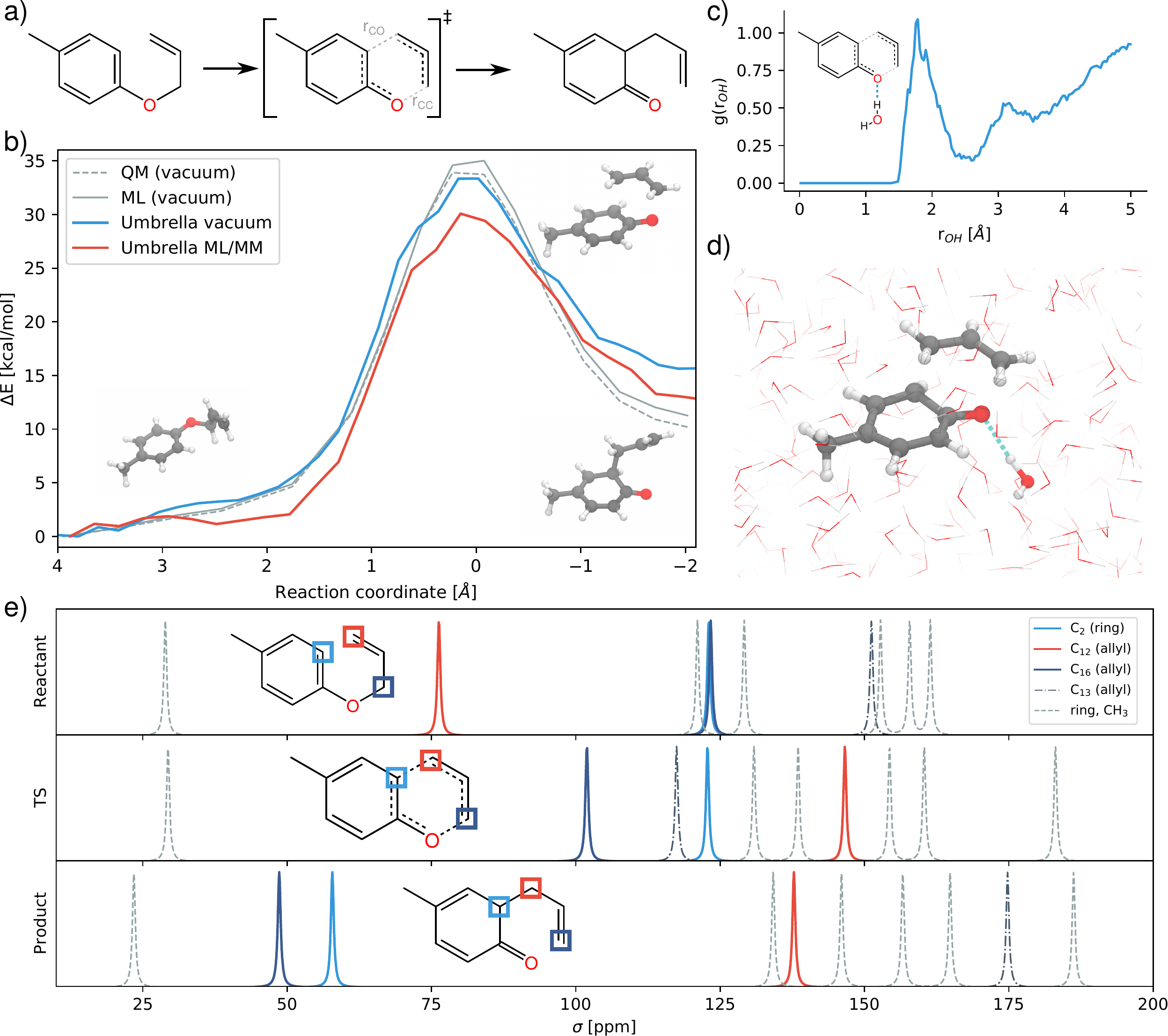}
	\caption{\textbf{Study of the allyl-p-tolyl ether Claisen rearrangement.} (a) Scheme of the reaction, with the two bonds used to determine the reaction coordinate marked in gray. (b) Reaction barriers as computed in gas-phase via nudged elastic band optimization (gray) and via umbrella sampling in vacuum (blue) and water (red). (c) Radial distribution function of the distance between ether oxygen and water hydrogens in the transition state. (d) Transition state configuration with stabilizing hydrogen bond. (e) NMR chemical shifts predicted for the different stages of the simulation.}
	\label{fig:ate}
\end{figure*}

An example for such a reaction is the Claisen rearrangement of allyl-p-tolyl ether (Fig.~\ref{fig:ate}a).
The presence of water as a solvent accelerates this reaction by a factor of 300 compared to reaction rates in the gas-phase~\cite{white1970claisen}.
Computational and experimental studies have determined that the main reason for this acceleration is explicit hydrogen bonding between the transition state and the water molecules of the solvent.
These lead to a lowered barrier and promote the reaction~\cite{irani2009joint, acevedo2010claisen}.
The combination of computational efficiency and accuracy with the ability to perform ML/MM simulations makes FieldSchNet well suited for modeling such reactions.
Beyond that, the model provides access to a range of properties which can be used to characterize the different species formed during reaction.

We train two FieldSchNet models to simulate the first step in the Claisen rearrangement of allyl-p-tolyl ether.
The first model is trained on PBE0 reference configurations sampled from a metadynamics trajectory of the reaction simulated at a lower level of theory.
A second model is a ML/MM model based on the reference configurations determined above augmented by different MM charge distributions sampled via the TIP3P force field for water~\cite{mackerell1998all} (see Sec~\ref{asec:reference} for details).
Errors for both models are reported in Supplementary Tab.~4.

We perform umbrella sampling~\cite{kastner2011umbrella} in vacuum as well as a solvent box containing $\sim$ 7000 TIP3P water molecules using the respective FieldSchNet models.
The difference between the two bonds broken and formed during rearrangement is chosen as reaction coordinate (indicated as $r_\mathrm{CO}$ and $r_\mathrm{CC}$ in Fig~\ref{fig:ate}a).
The speedup offered by FieldSchNet for this system is even more pronounced than for ethanol.
A single computation which takes 1.6~hours with the reference method can now be performed in 180~ms, corresponding to an acceleration by a factor of over $\sim$30\,000.

The resulting free energy barriers are depicted in Fig.~\ref{fig:ate}b, along with potential energy barriers computed in vacuum using the reference method and vacuum model.
The FieldSchNet ML/MM model correctly predicts a lower activation barrier (30.08~kcal/mol) for the aqueous environment compared to the gas phase reaction (33.35~kcal/mol).
The overall difference in the barrier height $\Delta\Delta G=3.28$~kcal/mol is close to the experimental value of $\Delta\Delta G=4$~kcal/mol.\cite{white1970claisen}
Analyzing the configurations sampled during the ML/MM simulation, we observe the hydrogen bonding between the ether oxygen and water molecules in the solvent responsible for the acceleration of the reaction~\cite{irani2009joint, acevedo2010claisen}.
Fig.~\ref{fig:ate}c shows the radial distribution function between hydrogens in the solvent and the oxygen of the transition state.
A pronounced peak at a distance of 2~\AA\ indicates that hydrogen bonds between solvent and solute form frequently at this stage.
Note that the absolute predicted height of the barriers is underestimated compared to experiment (38.4 and 34.4~kcal/mol for vacuum and water, respectively).
An analysis of the gas phase reaction barrier reveals that this is due to the reference and not an artifact of the machine learning model (Fig~\ref{fig:ate}b).

The NMR chemical shifts provided by FieldSchNet can be used to trace structural changes occurring during the rearrangement and allow to connect theoretical predictions to experiment.
Fig.~\ref{fig:ate}d depicts the $^{13}$C chemical shifts predicted for different stages of the reaction.
For example, it shows how the first (C$_{12}$) and third (C$_{16}$) carbon atom in the allyl ether exchange chemical environments during reaction.
The former moves from typical shifts of allyl ether groups ($\sim$68~ppm) to shifts associated with terminal carbons in conventional allyl groups ($\sim$130~ppm).
At the same time, the C$_{16}$ undergoes this process in reverse, ending at shifts of $\sim$50~ppm characteristic for this position in allyl substituents.
Another change of interest are the shifts of the carbon C$_2$ in the aromatic ring where a new bond forms.
This atom starts at typical values for aromatic carbons ($\sim$120~ppm), staying there during the transitions state.
Upon formation of the product, the aromaticity of the ring is lost and the shift moves to regions more indicative for carbon atoms in cyclic ketones ($\sim$25-50~ppm).

\subsection{Designing molecular environments}
The analytic nature of neural networks allows to establish a direct relation between molecular structure and properties which can be exploited in inverse chemical design applications~\cite{gebauer2019symmetry,schutt2019unifying}.
FieldSchNet is well suited for such tasks, as it provides access to a wide range of response properties as a function not only of molecular structure but also the external environment.
This offers the possibility to manipulate external fields and molecular environments in order to optimize certain properties or to control reaction rates.
In the following, we apply FieldSchNet to design a chemical environment promoting the Claisen rearrangement reaction studied above.

\begin{figure}
	\includegraphics[width=\columnwidth]{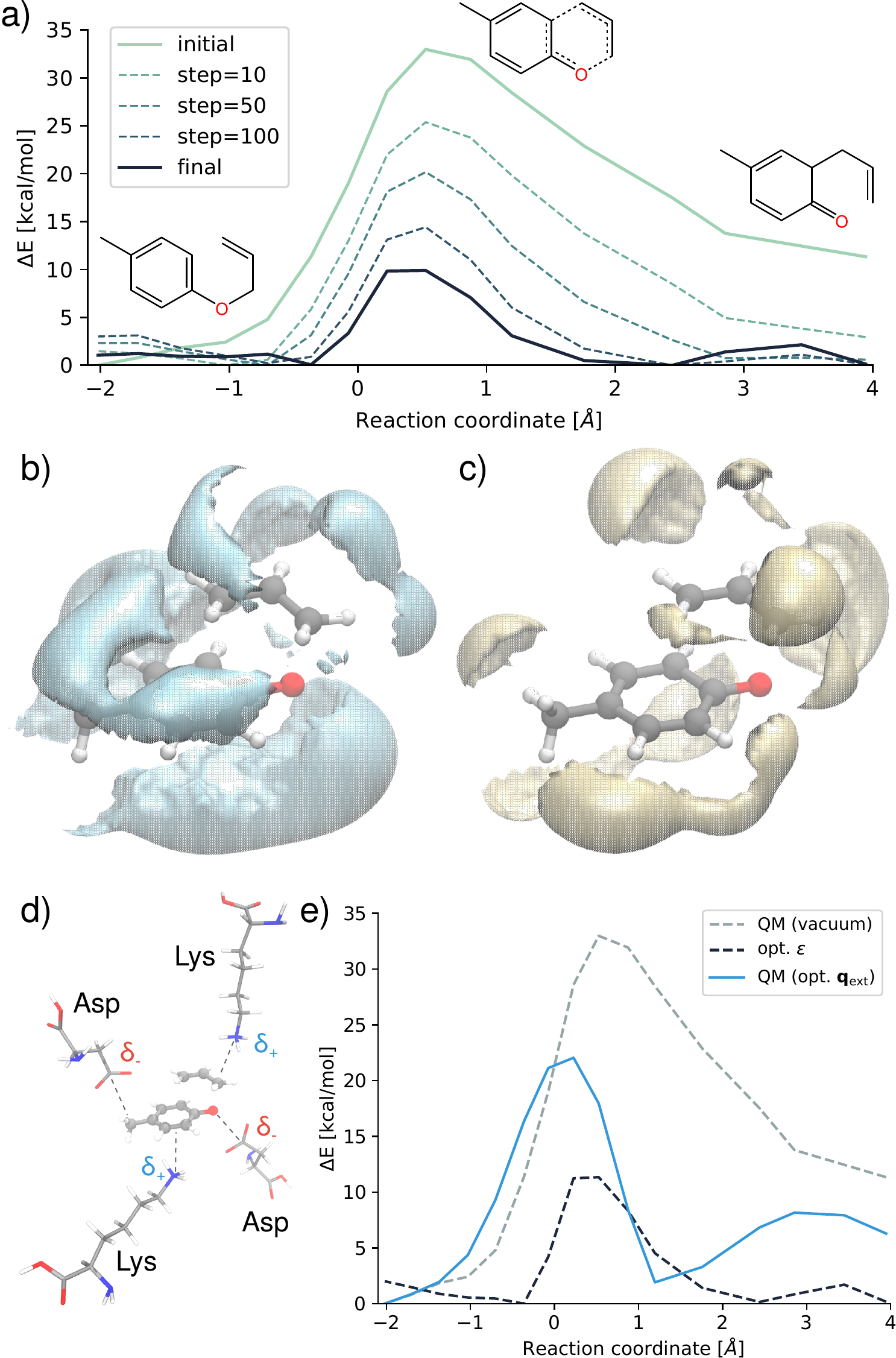}
	\caption{\textbf{Design of reaction environments.} (a) Evolution of the reaction barrier height during optimization of the environment. (b) Distribution of regions of negative charge around the transition state. (c) Regions of positive charge. (d) Transition state with optimal placement of charged amino acid residues (e) Reaction barrier recomputed for the optimal external charge arrangement generated by the amino acid residues (blue) compared to the original barrier (gray) and the barrier obtained for the optimized field (black).}
	\label{fig:field_opt}
\end{figure}
The external field in the ML/MM FieldSchNet model used to describe the reaction depends on the charge distribution of the environment.
Thus, the external charges can be optimized to lower the reaction barrier by minimizing
\begin{equation}
\mathcal{L} = \sum_j^{N_\mathrm{img}} \left( E_j( \mathbf{q}_\mathrm{ext} ) - \min[\{E_j(\mathbf{q}_\mathrm{ext})\}] \right)^2,
\label{eq:invdesign}
\end{equation}
where $j$ indicates the images along the reaction path obtained via nudged elastic band search and $\min[\{E_j(\mathbf{q}_\mathrm{ext})\}]$ is the minimal energy encountered along the path.
The external charges $\mathbf{q}_\mathrm{ext}$ are placed on a grid surrounding a cavity shaped by the reaction and are initialized to zero.

Fig.~\ref{fig:field_opt}a shows the evolution of the barrier during various stages of the optimization procedure.
By designing an optimal environment, the activation barrier can be reduced by $\sim25$~kcal/mol, lowering it from an initial 35~kcal/mol to 10~kcal/mol.
These findings correspond to a rate acceleration by a factor of $\sim2\times10^{18}$ at 300~K.
Fig.~\ref{fig:field_opt}b and c show the optimized environment in presence of the transition state, visualizing regions of negative (b) and positive (c) charge.
Motifs such as the strong negative density close to the oxygen atom and the neighboring carbons involved in the forming bond are in strong agreement with experimental studies, where it was found that electron donating groups in these regions promote the reaction~\cite{irani2009joint, coates1987synthesis}.

We outline how such an optimized field can guide design endeavors by translating these motifs into an atomic environment of amino acids as would be present in an enzymatic active site.
Negatively charged aspartic acid residues (Asp) are placed close to regions exhibiting high negative density, while regions of high positive charge are populated with lysine (Lys) molecules.
Using Eq.~\ref{eq:invdesign}, we optimize the placement of the amino acids based on the electrostatic field derived from their Hirshfeld atomic partial charges\cite{hirshfeld1977bonded}.
The resulting environment is shown in Fig~\ref{fig:field_opt}d.
Finally, we recompute the reaction barrier in presence of the amino acid charge distribution with the original electronic structure reference (Fig.~\ref{fig:field_opt}e).
Although the optimal environment cannot be reconstructed perfectly due to the constraint imposed by the structure of the molecules, this relatively straightforward approach leads to a significant reduction of the barrier.
The activation energy is lowered from 35~kcal/mol to 22~kcal/mol, which still corresponds to an acceleration by a factor of $\sim3\times10^{9}$.
This demonstrates that the theoretically optimized environment Fig.~\ref{fig:field_opt}b and d can be mimicked retaining the same trend on the barrier and reaction speed.

\section{Discussion}
FieldSchNet enables modelling the interactions of molecules with arbitrary external fields.
This offers access to a wealth of molecular response properties such as polarizabilities and nuclear shielding tensors without the need for introducing specialized models.
Leveraging external fields as an interface, the model can further operate as a polarizable continuum model for solvents, as well as a surrogate for the quantum mechanics region in quantum mechanics/molecular mechanics (QM/MM) schemes, yielding a ML/MM approach.
As a consequence, FieldSchNet paves the way to many promising applications out of reach for previous techniques.

Computational spectroscopy can benefit greatly from the FieldSchNet framework, as it provides efficient means for computing high quality spectra as we have demonstrated for IR and Raman spectra as well as NMR chemical shifts.
Combining the latter with nuclear spin-spin coupling tensors, i.e. response properties of the magnetic moments, enables the accurate prediction of nuclear magnetic resonance spectra.
In principle, all other spectroscopic quantities, derived via the response formalism, can be modeled by FieldSchNet as well.

The ability of our neural network to operate as a continuum model for solvation is not only highly attractive for applications in drug design, where efficient models of solvent effects are much sought after, but serves as a starting point for developing more powerful models, which could for example consider structural aspects of the solvent.
When treating interactions with the environment explicitly, the introduced ML/MM procedure proves to be a powerful tool combining the accuracy of machine learning potentials with the even higher speed of force fields.
While the presented study of solvent effects on the Claisen rearrangement reaction of allyl-p-tolyl ether would require 18~CPU years with the electronic structure reference, it was performed within 5~hours with FieldSchNet on a single GPU.
This greatly expands the range of application of ML models and brings  the simulation of large, biologically relevant systems, such as enzymatic reactions, within reach.

Moreover, the fully analytic nature of FieldSchNet enables inverse design as we have illustrated by minimizing the reaction barrier of the Claisen rearrangement through interactions with an optimized environment.
Coupling this to a generative model of molecular structure~\cite{gebauer2019symmetry}, the charge distributions found using FieldSchNet may be populated in a fully automated fashion.
Possible application of FieldSchNet to inverse design tasks include the targeted functionalization of compounds or the creation of enzyme cavities promoting reactions.

FieldSchNet constitutes a unified framework for describing reactions and spectroscopic properties in solution.
Beyond that, it provides insights on how these quantities can be controlled via the molecular environment.
As this opens up new avenues for designing workflows tightly integrated with experiment, we expect FieldSchNet to become a valuable tool for chemical research and discovery.

\clearpage

\section{Methods}

\subsection{Reference data generation}\label{asec:reference}

All electronic structure reference computations were carried out at the PBE0/def2-TZVP\cite{adamo1999toward,Weigend2005PCCP} level of theory using the ORCA quantum chemistry package~\cite{Neese2012WCMS}.
SCF convergence was set to tight and integration grid levels of 4 and 5 were employed during SCF iterations and the final computation of properties, respectively.
In the case of allyl-p-tolyl ether, the RIJK approximation was used to accelerate computations~\cite{weigend2002fully}.
Nuclear shielding tensors were computed with the Gauge Including Atomic Orbitals approach\cite{helgaker1999ab} implemented in ORCA, while continuum solvents calculations were performed with the in package conductor-like polarizable continuum model~\cite{barone1998quantum}.

The reference data for ethanol was generated by selecting 10\,000 random configurations from the MD17 database~\cite{chmiela2017machine} and recomputing them at the above level of theory.
In addition, continuum solvent calculations for the four studied continuum solvents (toluene, ethanol, methanol, water) were carried out for the structures selected in this manner.
A training set for continuum models containing 30\,000 ethanol configurations were constructed by merging the vacuum, ethanol and water data. 
Reference data for the ML/MM simulations was generated in a two-step approach. Initially, a periodic box of 1250 ethanol molecules was equilibrated with the NAMD molecular dynamics package~\cite{phillips2005scalable} using the CHARMM General Force Field~\cite{vanommeslaeghe2010charmm} for 1~$\mu$s.
Using the native NAMD interface to ORCA, electrostatic embedding QM/MM simulations were carried out, where one of the ethanols was described at the PBE/def2-SVP~\cite{perdew1996generalized, Weigend2005PCCP} level of theory. CHELPG charges~\cite{breneman1990determining} were used as partial charges for the quantum regions and simulations were run for 50~ps using 0.5~fs time steps.
For all simulations, temperatures were kept at 300~K using a Langevin thermostat~\cite{bussi2007langevin} and pressures at 1~atm using a Langevin piston barostat~\cite{feller1995langevin}.
From this trajectory, 30\,000 QM ethanol configurations and the associated charge distributions of the environment were sampled at random and recomputed at the PBE0/def2-TZVP level.

Reference data for the allyl-p-tolyl ether Claisen rearrangement reaction in vacuum was obtained via metadynamics~\cite{barducci2011metadyn} at the PBE/def2-SVP level of theory.
The two bonds involved in the reaction (see Fig.~\ref{fig:ate}) were selected as collective coordinates and Gaussians with a height of 1~kcal/mol and a width of 0.529~\AA\ were deposited each 100 simulation steps.
The system was simulated for a total of 50~ps using 0.5~fs time steps.
Temperature was kept constant at 500~K by means of a Nose-Hoover chain thermostat~\cite{martyna1992nhc}.
We then selected 61\,000 configurations from this and recalculated them with the reference level of theory.
Data for MM/ML simulations was generated by suspending 20334 configurations sampled during metadynamics in periodic solvation boxes with 9260 TIP3P waters~\cite{mackerell1998all}.
Keeping the allyl-p-tolyl ether coodinates frozen, the water box was then optimized and simulated for 50~ps with NAMD. 
For temperature and pressure control, the same setup as in the ethanol box was used.
From each of these boxes, 3 ether configurations and associated charge distributions were drawn and recomputed at the PBE0/def2-TZVP level, yielding 61\,002 reference data points.

\subsection{Training setting}
The training settings for each data set are provided in Supplementary Tab.~1.
The initial learning rates were decayed by a factor of 0.8 after $t_\mathrm{patience}$ epochs of no improvement.
Training was stopped the learning rate reached a value of 1e-6.
The dipole cutoffs were chosen to be the same as the cutoffs for the interactions.
Supplementary Tab.~2 provides the tradeoffs $eta$ used for the different response properties in the composited loss function minimized during training.
As the $^1$H, $^{13}$C and $^{17}$O chemical shifts lie on completely different scales, the shielding tensor loss terms for each atom were weighted by an element dependent factor in order to achieve equal relative accuracy between all contributions.
We used factors of $\omega_H = 1.0$, $\omega_C = 0.167$ and $\omega_O = 0.022$, which were determined based on the reference data.
The FieldSchNet model and training procedures were implemented using PyTorch~\cite{NEURIPS2019_9015} and the SchNetPack code package for machine learning in atomistic systems~\cite{schuett2018schnetpack}.

\subsection{Model for continuum solvation}\label{asec:pcm}

A FieldSchNet-based machine learning potential for continuum solvent effects can be derived by adapting the Onsager expression for the reactive field~\cite{onsager1936electric}.
Modeling the external electric field $\boldsymbol{\epsilon}_\textrm{solv}$ experienced by each atom $i$ due to a continuum solvent with dielectric constant $\varepsilon$ as
\begin{align}
\boldsymbol{\epsilon}_\text{solv}(\mathbf{R}_i)^l &= \frac{2(\varepsilon-1)}{(2\varepsilon+1)} \boldsymbol{\mu}_i^l  a_i^l\\
a_i^l &= \mlp\left(\mathbf{x}_i^l\right)
\end{align}
allows for a direct coupling between the molecular potential energy and the solvent.
The solvent field is adapted in each layer $l$ based on the atomic dipole features $\boldsymbol{\mu}_i^l$.
The term $a^l_i$ models the effective volume of the nucleus accessible due to its environment and is modeled by a neural network.

\subsection{Model for hybrid machine learning/molecular mechanics (ML/MM)}\label{asec:mlmm}
We adopt a QM/MM-like approach for FieldSchNet by coupling the classical and quantum region with electrostatic embedding, resulting in a ML/MM approach.
In this case, the quantum region is polarized by the charges of the classical region, while the electrostatic energy of the classical region is in turn modified by the point charges computed for the quantum region.
The influence of the external charges on the molecule can be modeled via the electrostatic field exerted by a collection of point charges
\begin{equation}
\boldsymbol{\epsilon}_\text{ext}\left(\mathbf{R}_i\right) = \sum_k \frac{q_k (\mathbf{R}_k - \mathbf{R}_{i})}{r_{ik}^5},
\end{equation}
where $q$ are the external charges and $\mathbf{R}_k$ the associated positions.
A suitable set of partial charges for the machine learning region can be obtained in the form of generalized atomic polar tensor charges~\cite{cioslowski1989general}, which are the response property
\begin{equation}
q_i = \frac{1}{3} \text{tr}\left(  \frac{\partial \boldsymbol{\mu}}{\partial \mathbf{R}_i}  \right) = -\frac{1}{3} \text{tr}\left(  \frac{\partial^2 E}{\partial \mathbf{R}_i \partial \boldsymbol{\epsilon}}  \right).
\end{equation}
These charges are fully polarized charges, depending not only on the molecular structure but also the charge distribution of the environment.

\subsection{Computational Details}
Unless stated otherwise, the velocity Verlet algorithm and a time step of 0.5~fs were used to integrate the equations of motion.
All simulations not using NAMD\cite{phillips2005scalable} were carried out with the molecular dynamics module implemented in SchNetPack~\cite{schuett2018schnetpack}.

Classical molecular dynamics simulations for ethanol in vacuum and continuum solvents were carried for 50~ps at a temperature of 300~K controlled via Nose-Hoover chain~\cite{martyna1992nhc} thermostat with a chain length of 3 and time constant of 100~fs. The first 10~ps of these trajectories were then discarded.
Ring polymer molecular dynamics were performed for 20~ps, using a time step of 0.2~fs and a specially adapted global Nose-Hoover chain as introduced in Ref.~\citenum{ceriotti2010stochastic} to keep the temperature at 300~K.
Once again, a chain length of 3 and time constant of 100~fs were chosen for the thermostat.

Simulations for the ethanol ML/MM model were carried out using a custom interface between NAMD and our machine learning code. First, a periodic box of 1250 ethanol molecules was equilibrated with NAMD for 1~$\mu$, using the CHARMM General Force Field~\cite{vanommeslaeghe2010charmm}.
Bonds to hydrogens were kept frozen with the SHAKE algorithm~\cite{RYCKAERT1977327} and a time step of 1~fs was used.
One ethanol was then selected for modeling via the FieldSchNet ML/MM model and ML/MM simulations were carried out using a custom interface between NAMD and our machine learning code for a total of 50~ps.
For both simulations, temperatures were kept at 300~K with a Langevin thermostat~\cite{bussi2007langevin} and pressures at 1~atm using Langevin piston barostat~\cite{feller1995langevin}.
The first 10~ps of the trajectory were discarded.

Umbrella sampling simulations for the Claisen rearrangement set up according to the following protocol.
Using the difference between the bonds formed and broken as the reaction coordinate, we determined the centers for the harmonic bias potentials by choosing 50 equidistant points along the reaction coordinate.
The centers ranged from values of -4.15~\AA\ to 5.18~\AA\, with an increment of 0.19~\AA.
For each center, we selected the closest lying structure in the metadynamics trajectory used for generating the reference data as a starting configuration for the umbrella sampling run.
All simulations used a force constant of 112.04~kcal/mol/\AA$^2$.
Umbrella sampling in vacuum was carried out for each window by first equilibrating the system for 25~ps using a Berendsen thermostat~\cite{Berendsen1984} at 300~K (time constant of 100~fs) followed by 25~ps production simulation at the same temperature with a Nose-Hoover chain (chain length of 3 and time constant of 100~fs).
For the ML/MM model umbrella simulations, the starting configurations were first solvated in a periodic box of 9260 water molecules treated with the TIP3P force field.
Keeping the allyl-p-tolyl ether structures frozen, the water box was first minimized and the equilibrated for 200~ps to a temperature of 300~K and pressure of 1~atm with a Langevin thermostat and Langevin piston barostat using NAMD.
Bonds involving water hydrogens were kept frozen with the SHAKE algorithm and a time step of 1~fs was used.
Starting from the systems prepared in this manner, ML/MM simulations were performed for 25~ps using the same pressure and temperature control as above.

Free energy profiles were constructed from the umbrella sampling data using the WHAM code with convergence set to 1e-9 and a temperature of 300~K~\cite{kumar1992wham, grossfieldwham}.

Infrared and polarized as well as depolarized Raman spectra were computed from the time-autocorrelation functions of the dipole moment and polarizability time derivatives according to the relations given in Ref.~\citenum{thomas2013computing}.
Autocorrelation functions were computed using the Wiener-Khinchin theorem~\cite{wiener1930} and a autocorrelation depth of 2048~fs.
In order to enhance the quality of the spectra, a Hann window function~\cite{6768513} and zero-padding were applied to the autocorrelation functions before computing the spectra.
A laser frequency of 514~nm and temperature of 300~K were used for calculating the Raman spectra.

NMR chemical shifts were computed as the average trace of the nuclear shielding tensor  $\sigma_i = \frac{1}{3}\mathrm{tr}[\boldsymbol{\sigma}_i]$.
These chemical shifts were then referenced to the shifts computed for a tetramethylsilane molecule via
\begin{equation}
\sigma_i = \sigma_\mathrm{ref}^{(Z)} - \sigma_i
\end{equation}
The reference shifts computed with the PBE0/def2-TZVP were $\sigma_\mathrm{H} = 31.77$~ppm and $\sigma_\mathrm{H} = 188.53$~ppm.

\clearpage

\subsection*{Data availability}
All datasets used in this work will be made available on http://www.quantum-machine.org/datasets after publishing.

\subsection*{Code availability}
All code developed in this work will be made available upon request.

\begin{acknowledgments}
This project has received funding from the European Unions Horizon 2020 research and innovation program under the Marie Sk\l{}odowska-Curie grant agreement No. 792572.
KTS and KRM acknowledge support by the Federal Ministry of Education and Research (BMBF) for the Berlin Center for Machine Learning / BIFOLD (01IS18037A).
KRM acknowledges financial support under the Grants 01IS14013A-E, 01GQ1115 and 01GQ0850; Deutsche Forschungsgemeinschaft (DFG) under Grant Math+, EXC 2046/1, Project ID 390685689 and KRM was partly supported by the Institute of Information \& Communications Technology Planning \& Evaluation (IITP) grants funded by the Korea Government(No. 2019-0-00079,  Artificial Intelligence Graduate School Program, Korea University).
Correspondence to MG.
\end{acknowledgments}

\subsection*{Author Contributions}
MG and KTS conceived the research. MG developed the method and carried out the reference computations and simulations.
KTS, MG, KRM designed the experiments and analyses. MG and KTS wrote the paper. KTS, MG and KRM discussed results and contributed to the final version of the manuscript.

\subsection*{Competing Interests}

The authors declare no competing interests.

\bibliographystyle{naturemag}
\bibliography{gastegger_field_schnet.bib}

\renewcommand{\thefigure}{\arabic{figure}}
\renewcommand{\thetable}{\arabic{table}}
\renewcommand{\figurename}{Supplementary Figure}
\renewcommand{\tablename}{Supplementary Table}

\captionsetup[figure]{justification=raggedright}
\captionsetup[table]{justification=raggedright}

\clearpage

\appendix

\section{Supplementary Information: Machine learning of solvent effects on molecular spectra and reactions}
	
\subsection*{Supplementary text 1: Response properties}\label{asec:response} 
	The energy predicted by FieldSchNet is an analytic function of the coordinates $\mathbf{R}$, as well as the external fields and their associated atomic dipole moments.
	This makes it possible to access so-called response properties, which are partial derivatives of the potential energy~\cite{jensen2007introduction}.
	Assuming the presence of an external electric $\boldsymbol{\epsilon}$ field and a magnetic field $\mathbf{B}$ with its corresponding nuclear magnetic moments $\{\mathbf{I}_i\}$, a general response property $\boldsymbol{\Pi}$ takes the form
	\begin{equation}
	\boldsymbol{\Pi}(n_\mathbf{R}, n_{\boldsymbol{\epsilon}}, n_\mathbf{B}, n_{\mathbf{I}_i}) = \frac{\partial^{n_\mathbf{R} + n_{\boldsymbol{\epsilon}} + n_\mathbf{B} + n_{\mathbf{I}_i}} E(\mathbf{R}, \boldsymbol{\epsilon}, \mathbf{B}, \mathbf{I}_i)}{\partial \mathbf{R}^{n_\mathbf{R}} \partial \boldsymbol{\epsilon}^{n_{\boldsymbol{\epsilon}}} \partial \mathbf{B}^{n_{\mathbf{B}}} \partial \mathbf{I}_i^{n_{\mathbf{I}_i}}},\label{eq:response}
	\end{equation}
	where the $n$s indicate the $n$-th order partial derivative w.r.t. the quantity in the subscript.
	A response property modeled by most machine learning potentials are the nuclear forces $\mathbf{F} = -\boldsymbol{\Pi}(1,0,0,0)$, which are the negative first derivative of the energy with respect to the nuclear positions.
	
	However, the expression above offers instructions on obtaining a wealth of other quantities, some of which are highly relevant for molecular spectroscopy and/or provide a direct connection to experiment.
	Infrared spectra can e.g. be simulated based on dipole moments $\boldsymbol{\mu} = -\boldsymbol{\Pi}(0,1,0,0)$, while
	molecular polariziabilities $\boldsymbol{\alpha} = -\boldsymbol{\Pi}(0,2,0,0)$ offer access to polarized and depolarized Raman spectra.
	A central response property of the magnetic field are nuclear magnetic shielding tensors $\boldsymbol{\sigma}_i = \boldsymbol{\Pi}(0,0,1,1)$.
	These allow the computation of chemical shifts recorded in nuclear magnetic resonance spectroscopy NMR via their average trace $\sigma_i = \frac{1}{3}\mathrm{tr}[\boldsymbol{\sigma}_i]$.
	
	The power of FieldSchNet (and field-based models in general) lies in the fact, that a single energy function provides access to a wide range of quantum chemical properties in a highly systematic manner.
	Moreover, the expression in Eq.~\ref{eq:response} above guarantees the correct geometric transformations of the property tensors with respect to rotations and translations of the molecule in the external field without the need of explicitly encoding the corresponding symmetries. 
	As is the practice with molecular forces, response properties can also be incorporated during training of the FieldSchNet model by including the appropriate squared errors into the loss function
	\begin{equation}
	\mathcal{L} = \eta_E (\tilde{E}-E)^2 +  \frac{1}{3N} \sum_{i}^N |\tilde{\mathbf{F}_i}-\mathbf{F}_i|^2 + \eta_{\boldsymbol{\mu}} \frac{1}{3} |\tilde{\boldsymbol{\mu}}-\boldsymbol{\mu}|^2 + \ldots
	\end{equation}
	Here, the trade-offs $\eta$ weight the importance of a property in the loss and $N$ is the total number of atoms. The properties predicted by FieldSchNet according to Eq.~\ref{eq:response} are indicated with a tilde.

	\begin{table*}
		\caption{Training parameters for all models.}
		\label{stab:training}
		\begin{tabular}{lrrrrrrrrr}
			\toprule
			Dataset & $n_\mathrm{train}$ & $n_\mathrm{valid}$ & $n_\mathrm{test}$ & $n_\mathrm{batch}$ & $\mathrm{lr}_\mathrm{init}$ & $t_\mathrm{patience}$ & $n_\mathrm{features}$ & $n_\mathrm{interactions}$ & $r_\mathrm{cutoff}$ [\AA]\\
			\midrule
			ethanol (vacuum)    & 8000    & 1000 &    1000 & 20 & 1e-4 & 15 & 256 & 6 & 5.0\\
			ethanol (continuum) & 16\,000 & 2000 &    9990 & 20 & 1e-4 & 15 & 256 & 6 & 5.0\\
			ethanol (ML/MM)     & 18\,000 & 2000 & 10\,000 & 20 & 1e-4 & 15 & 256 & 6 & 5.0\\
			ether (vacuum)      & 50\,000 & 5000 &    6000 & 10 & 1e-4 & 25 & 256 & 5 & 5.0\\
			ether (ML/MM)       & 50\,000 & 5000 &    6002 & 10 & 1e-4 & 25 & 256 & 5 & 5.0\\
			\bottomrule
		\end{tabular}
	\end{table*}
	
	\begin{table}
		\caption{Tradeoffs $\eta$ used for training the different properties, assuming all quantities use atomic units.}
		\label{stab:tradeoffs}
		\begin{tabular}{lrrrr}
			\toprule
			&   ethanol  && \multicolumn{2}{c}{allyl-p-tolyl ether} \\
			\cmidrule{2-2} \cmidrule{4-5}
			Property                           &  all  && vacuum & ML/MM   \\
			\midrule
			$\mathbf{E}$                       & 1.0   &&   1.0   &    1.0 \\ 
			$\mathbf{F}$                       & 10.0  &&   5.0   &    5.0 \\ 
			$\boldsymbol{\mu}$                 & 0.01  &&   0.05  &    0.01 \\ 
			$\boldsymbol{\alpha}$              & 0.01  &&   0.001 &    0.0001 \\ 
			$\boldsymbol{\sigma}_\mathrm{all}$ & 0.05  &&   10.0  &    0.1 \\ 
			\bottomrule
		\end{tabular}
	\end{table}

	\begin{table*}
		\caption{\textbf{Test set performance of ethanol models.} Mean absolute errors of FieldSchNet trained on ethanol in vaccuum and pc-FieldSchNet trained with vaccum, ethanol and water as solvents. Solvents marked by * have not been used to train the continuum model.}
		\label{stab:ethanol}
		\begin{tabular}{llrrrrrrrrrr}
			\toprule
			\textbf{Property} &            \textbf{Unit}                   &    \textbf{Vacuum}     & \hspace{.3cm} & \multicolumn{5}{c}{\textbf{Continuum}}                                   & \hspace{.3cm} &   \textbf{ML/MM}        \\
			\cmidrule{5-9}	
			&                           &   &&   \textit{vacuum} & \textit{toluene}* & \textit{ethanol} & \textit{methanol}*&  \textit{water}  &&   \\
			\midrule
			E                              & kcal\,mol$^{-1}$              &   0.017 &&   0.035 &   0.137 &   0.052 &   0.056 &   0.062 &&   0.557 \\ 
			\textbf{F}                     & kcal\,mol$^{-1}$\,$\AA^{-1}$  &   0.128 &&   0.145 &   0.174 &   0.139 &   0.140 &   0.142 &&   0.683 \\ 
			$\boldsymbol{\mu}$             & D                             &   0.004 &&   0.004 &   0.006 &   0.005 &   0.005 &   0.005 &&   0.007 \\ 
			$\boldsymbol{\alpha}$          & Bohr$^3$                      &   0.008 &&   0.007 &   0.243 &   0.007 &   0.007 &   0.008 &&   0.010 \\ 
			$\boldsymbol{\sigma}_\mathrm{all}$ & ppm                       &   0.169 &&   0.157 &   0.149 &   0.140 &   0.140 &   0.141 &&   0.154 \\ 
			$\sigma_{H}$                   & ppm                           &   0.123 &&   0.122 &   0.116 &   0.113 &   0.113 &   0.114 &&   0.094 \\ 
			$\sigma_{C}$                   & ppm                           &   0.194 &&   0.186 &   0.175 &   0.166 &   0.166 &   0.167 &&   0.182 \\ 
			$\sigma_{O}$                   & ppm                           &   0.401 &&   0.312 &   0.298 &   0.248 &   0.248 &   0.250 &&   0.453 \\
			\bottomrule
		\end{tabular}
	\end{table*}
	
	\begin{table}
		\caption{Test set errors obtained for the allyl-p-toly Claisen rearrangement datasets.}
		\label{stab:ether}
		\begin{tabular}{llrrr}
			\toprule
			\textbf{Property}                       & \textbf{Unit}                          & \textbf{Vacuum}  && \textbf{ML/MM}   \\
			\midrule
			$\mathbf{E}$                              & kcal\,mol$^{-1}$              &   0.084 &&    0.400 \\ 
			$\mathbf{F}$                              & kcal\,mol$^{-1}$\,$\AA^{-1}$  &   0.141 &&    0.454 \\ 
			$\boldsymbol{\mu}$                         & D                             &   0.003 &&    0.026 \\ 
			$\boldsymbol{\alpha}$                       & Bohr$^3$                      &   0.039 &&    0.157 \\ 
			$\boldsymbol{\sigma}_\mathrm{all}$          & ppm                           &   0.273 &&    1.144 \\ 
			$\sigma_{H}$                   & ppm                           &   0.045 &&   0.154 \\ 
			$\sigma_{C}$                   & ppm                           &   0.301 &&  1.331 \\ 
			$\sigma_{O}$                   & ppm                           &   2.732 &&   11.144 \\ 
			
			\bottomrule
		\end{tabular}
	\end{table}

\end{document}